\def\bq{\begin{equation}}
\def\eq{\end{equation}}
\def\bqa{\begin{eqnarray}}
\def\eqa{\end{eqnarray}}
\def\bqb{\begin{eqnarray*}}
\def\eqb{\end{eqnarray*}}
\documentstyle[12pt]{article}\textwidth 16cm
\def\pr#1#2#3{ Phys. Rev. ${\bf{#1}}$ (#2) #3}
\def\prl#1#2#3{ Phys. Rev. Lett. ${\bf{#1}}$ (#2) #3}
\def\pl#1#2#3{ Phys. Lett. ${\bf{#1}}$ (#2) #3 }

\def\np#1#2#3{ Nucl. Phys. ${\bf{#1}}$ (#2) #3}
\def\zp#1#2#3{ Z. Phys. ${\bf{#1}}$ (#2) #3}

\global\nulldelimiterspace = 0pt
 
\def\Bsl{\hbox{/\kern-.6700em$B$}} 
\def\Dsl{\hbox{/\kern-.6700em$D$}} 
\def\Wsl{\hbox{/\kern-.6700em$W$}} 

\def\roughly#1{\mathrel{\raise.3ex
    \hbox{$#1$\kern-.75em\lower1ex\hbox{$\sim$}}}}

\def\mh2{m^2_H}

\begin{document}


\font\fifteen=cmbx10 at 15pt
\font\twelve=cmbx10 at 12pt

\begin{titlepage}

\begin{center}

\renewcommand{\thefootnote}{\fnsymbol{footnote}}

{\twelve Centre de Physique Th\'eorique\footnote{
Unit\'e Propre de Recherche 7061
}, CNRS Luminy, Case 907}

{\twelve F-13288 Marseille -- Cedex 9}

\vspace{3 cm}

{\fifteen THEORETICAL OVERVIEW ON DIBOSON PRODUCTION\footnote{Invited talk
given at $XI^{th}$ Topical Workshop on Proton Antiproton Collider Physics,
26 May-1 June 1996, Abano-Terme (Italy).}}

\vspace{0.3 cm}

\setcounter{footnote}{0}
\renewcommand{\thefootnote}{\arabic{footnote}}

{\bf 
Pierre CHIAPPETTA 
}

\vspace{2,3 cm}

{\bf Abstract}

\end{center}

Precise measurements of weak vector bosons self couplings give a hint 
on the electroweak symmetry breaking sector. We first stress that
present data from LEP and TEVATRON clearly indicate that weak bosons
are self interacting. We then review the limits on the
trilinear and quadrilinear couplings expected at LEP2, $e^+e^-$ linear
colliders and LHC.

\vspace{2,3 cm}


\bigskip

\renewcommand{\thefootnote}{\fnsymbol{footnote}}

\noindent Number of figures: 3

\bigskip

\noindent July 1996

\noindent CPT-96/P.3365

\bigskip

\noindent anonymous ftp or gopher: cpt.univ-mrs.fr

\end{titlepage}

\setcounter{footnote}{0}
\renewcommand{\thefootnote}{\arabic{footnote}}


Electroweak gauge bosons are well suited to test in a subtle way two
fundamental principles: gauge invariance and electroweak symmetry
breaking mechanism- hereafter denoted as EWSB. The first one gives the
strength of the trilinear and quadrilinear gauge bosons couplings -
whose existence is due to the non abelian structure of the gauge
group- whereas the longitudinal components of W and Z bosons directly
probe the EWSB mechanism (linear or non linear realization?).
 
\section{Restrictions from present data}
Since the standard model- hereafter denoted as SM- has been tested
at one loop level, we have now evidence that gauge bosons are self
interacting. First evidence came already in 1994 when it was shown     
\cite{GS94} that the $sin^2 \theta_W (M_Z)$  value predicted
without taking into account bosonic loops deviates from data by $7
\sigma$. As of today, the  three dimensional plot shown in
figure 1 clearly indicates the need for bosonic loops to match the SM
with precise data on $\frac{M_W}{M_Z}$,  $sin^2 \theta_W (M_Z)$
and the leptonic width $\Gamma_l$ \cite{DSW95}. 
	There are seven independent $ZWW$ form factors and six $\gamma WW$
besides the electric charge. Restricting our analysis to C and P
conserving $VWW$ couplings we parametrize the lagrangian as
\cite{HPZH}:

 \bqa && L_{VWW} = -ie \lbrack A_{\mu} (W^{- \mu \nu}
W^+_{\nu} -  W^{+ \mu \nu}W^-_{\nu}) + (1+ \Delta \kappa_{\gamma})
F_{\mu \nu}W^+_{\mu} W^-_{\nu} \nonumber\\ 
&& + \cot\theta_W (1+ \Delta g_1^Z) Z_{\mu} (W^{- \mu \nu}
W^+_{\nu} - W^{+ \mu \nu}W^-_{\nu})  + (1+ \Delta \kappa_Z)
Z_{\mu \nu}W^+_{\mu} W^-_{\nu} \nonumber\\ 
&& + \frac{1}{M_W^2} (\lambda_{\gamma} F^{\nu \lambda}+ \cot\theta_W
\lambda_Z Z^{\nu \lambda}) W^+_{\lambda \mu} W^{- \mu}_{\nu}
 \rbrack  \label{param} \eqa

  In the SM $\Delta g_1^Z=0$, $\Delta \kappa_V = 0$ and
$\lambda_V = 0$, whereas no self interaction among gauge bosons would
lead to   
 $\Delta \kappa_V = - 1$ and $\lambda_V =0$.
Tevatron collider \cite{CDFD0}, assuming $\lambda_{\gamma}= \lambda_Z$
and $\kappa_{\gamma}= \kappa_Z$, has excluded  $\Delta \kappa_V = -1$ 
since $ -0.7 \leq \Delta \kappa \leq 0.89$ and $-0.44 \leq \lambda \leq
0.44$. 
For the moment the sensitivity to the Higgs mass is weak: $65.2 GeV
\leq M_H \leq 440 GeV$ \cite{HOLLIK}, and the preference for small
Higgs masses rely entirely on observables which differ from SM
expectations by $2-3 \sigma$ i.e. $R_b = \frac{\Gamma(Z \rightarrow b
\bar b)}{\Gamma_{hadronic}}$,   $R_c = \frac{\Gamma(Z \rightarrow c
\bar c)}{\Gamma_{hadronic}}$ and the left right asymmetry $A_{LR}$.

In order to probe the EWSB a precise measurement of trilinear and
quadrilinear gauge boson couplings is mandatory. For this purpose we
will introduce the notion of effective lagrangian, which gives a
general description of the phenomenon without knowing precisely its
origin or the underlying theory\cite{ALVARO}, \cite{H93}. The inputs are
the known symmetries at low energies and the particle content. At the
electroweak scale we have to keep $SU(2) \times U(1)$ invariance and
the custodial $SU(2)$ symmetry- since $\Delta \rho \leq 4 10^{-3}$
indicates that weak isospin breaking effects are small. The residual
interactions affecting the self couplings are described by operators
$O_i$: \bq L_{eff} = \sum_{n} \sum_i \frac{f_i^{(n)}}{\Lambda^n}
O_i^{(n+4)} \eq  
where $\Lambda$ is the scale for new physics. Since we consider
low energies (smaller than $\Lambda$) we will restrict our analysis to
dimension 6 operators. The introduction of anomalous couplings leads
to a violation of unitarity which is cured either by introducing form
factors or imposing unitarity constraints on the couplings.

Some of these operators are already constrained at LEP1\cite{epsilon},
 since they
affect gauge boson two point functions. We shall first discuss the
scenario of a linear realization of EWSB through a Higgs doublet
$\Phi$. Precisely the operator $O_{\Phi,1} = {(D_{\mu} \Phi)}^+ \Phi
{\Phi}^+ (D^{\mu} \Phi)$ is restricted by the parameter $\epsilon_1$
\cite{LEP1} whereas the variable $\epsilon_3$  restricts $O_{BW} =
\frac{g g^{\prime}}{4} \Phi^+ B_{\mu \nu} W^{\mu \nu} \Phi$ (B being the
$U(1)$ field and W the $SU(2)$ field). The coefficient $f_{\Phi,1}$ is
of the order of $10^{-1}$ whereas $f_{BW}$ is of the order of 1
\cite{H93}. The operators describing the self interactions which do not
contribute to the two point functions are:
 \bq O_{W} = \frac{ig}{2} (D_{\mu} \Phi)^+\vec{\tau}.\vec W^{\mu \nu}
D_{\nu} \Phi \eq 
 
\bq O_{B} = \frac{ig^{\prime}}{2} (D_{\mu} \Phi)^+ B^{\mu \nu}
D_{\nu} \Phi \eq
and 
 \bq O_{WWW} = Tr \lbrack \frac{ig}{2}\vec{\tau}.\vec W_{\mu \nu}
\frac{ig}{2}\vec{\tau}.\vec W^{\nu \rho}
 \frac{ig}{2}\vec{\tau}.\vec W^{\mu}_{\rho} \rbrack 
 \eq . 
 Their contribution to the parameters\cite{HPZH} of $L_{VWW}$ given in
eq.\ref{param}
 reads:

\bq \Delta \kappa_{\gamma}=(f_B+f_W) \frac{M^2_W}{2 \Lambda^2} \eq
\bq \Delta g_1^Z= f_W \frac{M^2_Z}{2 \Lambda^2} \eq
\bq \Delta \kappa_Z= \lbrack f_W- \sin^2 \theta_W (f_B+f_W) \rbrack 
\frac{M^2_Z}{2 \Lambda^2} \eq

\bq \lambda_{\gamma} = \lambda_Z = f_{WWW} \frac{3 g^2 M^2_W}{2
\Lambda^2} \eq

The present limits on the coefficients of these operators are: 
\bq f_i \frac{v^2}{\Lambda^2} \simeq 10 \eq

We shall see later how future colliders will improve this sensitivity.
Let us now consider the non linear realization of EWSB where the Higgs
part of the lagrangian is replaced by:

\bq L_{EWSB} = \frac{v^2}{4} Tr (D^{\mu}\Sigma^+ D_{\mu} \Sigma )\eq 

where 
$\Sigma =  \exp(i \frac{\vec \tau.\vec{\pi}}{v})$.
LEP1, through $\epsilon_3$, constrains the operator: 
\bq O_{10} =\frac{g g^{\prime}}{16 \pi^2} Tr (B_{\mu \nu}\Sigma^+ W^{\mu
\nu} \Sigma) \eq.
The coefficient $L_{10}$ lies\cite{LANGACKER} in the range $-0.7 \leq
L_{10} \leq 2.4$. 
Instead of $O_W$ and $O_B$ we have now to consider the operators: 

\bq O_{9R}= - \frac{ig^{\prime}}{16 \pi^2} Tr(B^{\mu \nu}
D_{\mu}\Sigma^+ D_{\nu}\Sigma)  \eq

and 
\bq O_{9L}= - \frac{ig}{16 \pi^2} Tr(W^{\mu \nu}
D_{\mu}\Sigma^+ D_{\nu}\Sigma)  \eq.
Their contribution to the trilinear gauge couplings parametrized by
$L_{VWW}$ given in eq.\ref{param} according to \cite{HPZH} reads:

\bq \Delta g_1^Z =  L_{9L} (\frac{e^2}{\sin^2 \theta_W}) 
(\frac{1}{32 \pi^2 \cos^2 \theta_W}) \eq

\bq \Delta \kappa_Z =  (L_{9L}+L_{9R}) (\frac{e^2}{\sin^2 \theta_W}) 
(\frac{1}{32 \pi^2}) \eq
 
Present limits are weak: $L_9 \simeq 10^3$.

\section{Prospects at future colliders.}

LEP2 is now starting to operate and will probe directly the self
interactions of electroweak gauge bosons. The most interesting
reaction is $e^+e^- \rightarrow W^+W^-$. The calculations performed by
the LEP2 working group \cite{LEP2} have taken into account initial
state radiation, W width effects and the background from four fermion
final state, since the optimal deaay channel is $jjl\nu$. At
$\sqrt S=190 GeV$ and for an integrated luminosity of $500 pb^{-1}$
limits on $L_9$ have improved:
$-30 \leq L_{9L} \leq 30$  and    $-300 \leq L_{9R} \leq 750$.
A more energetic $e^+e^-$ linear collider, like a NLC operating at 
$\sqrt S=500 GeV$ and for an integrated luminosity of $10 fb^{-1}$
will drastically constrain the parameters. The mode $\gamma \gamma
\rightarrow W^+W^-$ helps to constrain $L_{9R}$. This is explicitely
shown in figure 2 from \cite{BOUDJEMA}.

Moving from $500 GeV$ to the TeV range allows to gain one order of
magnitude. The best limits at $\sqrt S = 1.5 TeV$ with an integrated
luminosity of $190 fb^{-1}$ are obtained by keeping all resonant
diagrams from the semi leptonic final state \cite{COUTURE}. As shown in
figure 3, the sensitivity reached is much better than the one expected
at LHC where the mode $pp \rightarrow WZ$ restricts $L_{9R}$ in the
range:  $-2 \leq L_{9R} \leq 3$.

A comment is is order now concerning hadronic colliders. It has
recently been shown \cite{BAUR}, \cite{OHNEMUS} that QCD corrections
may be huge. This is the case for $W \gamma$ final state whose
next-to-leading correction increases the Born prediction from $20
\%$ at $\sqrt S= 2 TeV$ up to $300 \%$ at $\sqrt S = 40
TeV$. This huge effect affects the two body cross section characterized
by a radiation amplitude zero(i.e. an exact amplitude zero for some
values of the scattering angle), for which the $2 \rightarrow 3$
subprocesses fill the dip in the $\gamma$ rapidity distribution at LHC.
The ZZ, $W^+W^-$ and WZ final states are also affected. The approximate
amplitude zero in the WZ final state suppresses the Born cross section
and therefore NLO corrections are larger than  for ZZ or WW processes. 
Collinear splittings like $qg \rightarrow Zq$ followed by $q \rightarrow
q^{\prime}W$ induce an increase of the order of : 

\bq \frac{g^2}{4 \pi \sin^2 \theta_W} ln^2 (\frac{P_T^2}{M_W^2}) \eq

(where $P_T$ is the gauge boson transverse momentum) precisely in the
large $P_T$ range sensitive to anomalous trilinear gauge bosons
couplings. The way to solve this problem is to cut on the extra jet,
leading to the definition of a $WW/WZ + 0 jet$ cross section,
increasing the Born cross section by at most $20 \%$. 
Let me stress that the reaction $e^+e^- \rightarrow W^+_L W^-_L$ (resp.
$W^+_L W^-_T$) probes $\Delta \kappa_V$ (resp. $g_1^Z$), whereas  $p
\bar p \rightarrow W^{\pm}_L Z_L$ probes $g_1^Z$. Transversely
polarized gauge bosons test $\lambda_V$. The advantage of reactions
like $p \bar p \rightarrow W\gamma$ or $e^+e^- \rightarrow \gamma \nu
\bar \nu, Z \nu \bar \nu$  \cite{TIKONOV} is that they probe
independently $\gamma WW$ and $ZWW$ trilinear couplings.

We shall finally focus on some specific models. The first one is
supersymmetry, hereafter denoted as SUSY. Trilinear gauge boson
couplings are sensitive to SUSY at one loop level. In order to get a
gauge independent and finite result satisfying unitarity one has to
add contributions from boxes having a vector like structure: this the
pinch technique \cite{PINCH}. These contributions have to be compared
to the SM ones, which are sensitive to top and Higgs masses. The SM
one loop corrections are small:
\bq \Delta \kappa_\gamma \sim 5 10^{-3}, \Delta \kappa_Z \sim 3
10^{-3}, \lambda_V \sim 10^{-3} \eq
The principal source for deviations arises from neutralinos and
charginos\cite{ARGYRES}. The most favourable situation occurs when
$M_{\frac{1}{2}} < m_0, A_0$, where  $M_{\frac{1}{2}}$(resp. $m_0$) is
the common gaugino(resp. scalar) mass at the GUT scale whereas $A_0$ is
the trilinear soft breaking term. One can reach $\Delta \kappa_V \sim
10^{-2}$.

	We shall discuss now technicolor models: only naive QCD scaled
versions are ruled out by present data. Due to the heavy top mass and
precise LEP1/SLC data two attractive models have recently emerged:
topcolor assisted technicolor\cite{TOPCOLOR} and the non commuting
extended technicolor \cite{SCT95}. The first one differs from SM by a
new interaction of type $SU(3) \times U(1)$ leading to colorons which
affect top production, and to an extra Z which may affect diboson
production if it couples strongly to light fermions\cite{KOMINIS}. The
non commuting extended technicolor model is based on the assumption that
the ETC group does not commute with $SU(2)_W$, this can be realized for
example by assigning the technifermions in a right handed doublet.
This model can explain the $R_b$ deviations from SM.
All technicolor models predict the existence of a rich spectrum of new
particles like pseudogoldstone bosons, technirho, technieta... Their
existence may affect diboson production\cite{LANE}. As an example
production of pairs of longitudinally polarized Z bosons from gluon
gluon subprocess could be strongly enhanced in the mass range above
the colored pseudogoldstone boson threshold\cite{LEE}. Nevertheless a
quantitative study of the phenomenological consequences for diboson
production from those viable models is lacking.

\section{Quartic couplings.}
Quartic couplings which violate $SU(2)$ custodial symmetry have
already been strongly constrained at LEP1 at the level of $10^{-2}$
and a linear $e^+e^-$ collider will not improve these
limits\cite{BRUNSTEIN}. At LHC a gain in sensitivity of roughly one
order of magnitude is expected from the other operators i.e.:
 
$\alpha_1 \lbrack Tr(V_{\mu} V_{\nu}) \rbrack ^2$ and 
$\alpha_2 (\lbrack Tr(V_{\mu} V_{\mu}) \rbrack ^2$ , 

where 
$V_{\mu}= (D_{\mu} \Sigma) \Sigma^+$.

 \section{Conclusions}

To conclude, we have already evidence that electroweak gauge bosons
are self interacting. The model independent way to parametrize the
self couplings among gauge bosons is to use effective lagrangians. To
constrain operators contributing to trilinear couplings at the same
level as those contributing to two point functions are already
constrained by LEP, an $e^+e^-$ linear collider at $\sqrt S = 500 GeV$
with an integrated luminosity of $50 fb^{-1}$ is ideal and should be
more efficient than LHC. LEP limits on quartic couplings will only be
improved by LHC.
 \vspace{0.5cm}

{Acknowledgements}
\par
This work has been partially supported by the EC contract
CHRX-CT94-0579.
I would like to thank F. Boudjema and J. Papavassiliou for enlightening
discussions during the preparation of this report. I wish to thank
the organizing comitee for the invitation and the pleasant stay
in  Abano.

\centerline {\ {\bf Figure Captions }}

Fig.1 Need for bosonic loops from present electroweak data  \cite{DSW95}. 
The ball is the $68 \%$ CL of data. The net that the ball hints is the full
SM prediction with a top mass varying from $100 GeV$ by steps of $20 GeV$
and a Higgs mass varying from $100$ GeV to $1$ TeV(from left to right). The
line with cubes corresponds to the purely fermionic contribution.\\

 Fig.2 Expected bounds on $L_{9L}$ and $L_{9R}$ from LEP2 and a
linear collider at $\sqrt S = 500 GeV$ \cite{BOUDJEMA}.\\
 
Fig.3  Expected bounds on $L_{9L}$ and $L_{9R}$ from 
linear colliders at $\sqrt S = 500 GeV$ and  at $\sqrt S = 1.5 TeV$
compared to LHC \cite{BOUDJEMA}.\\

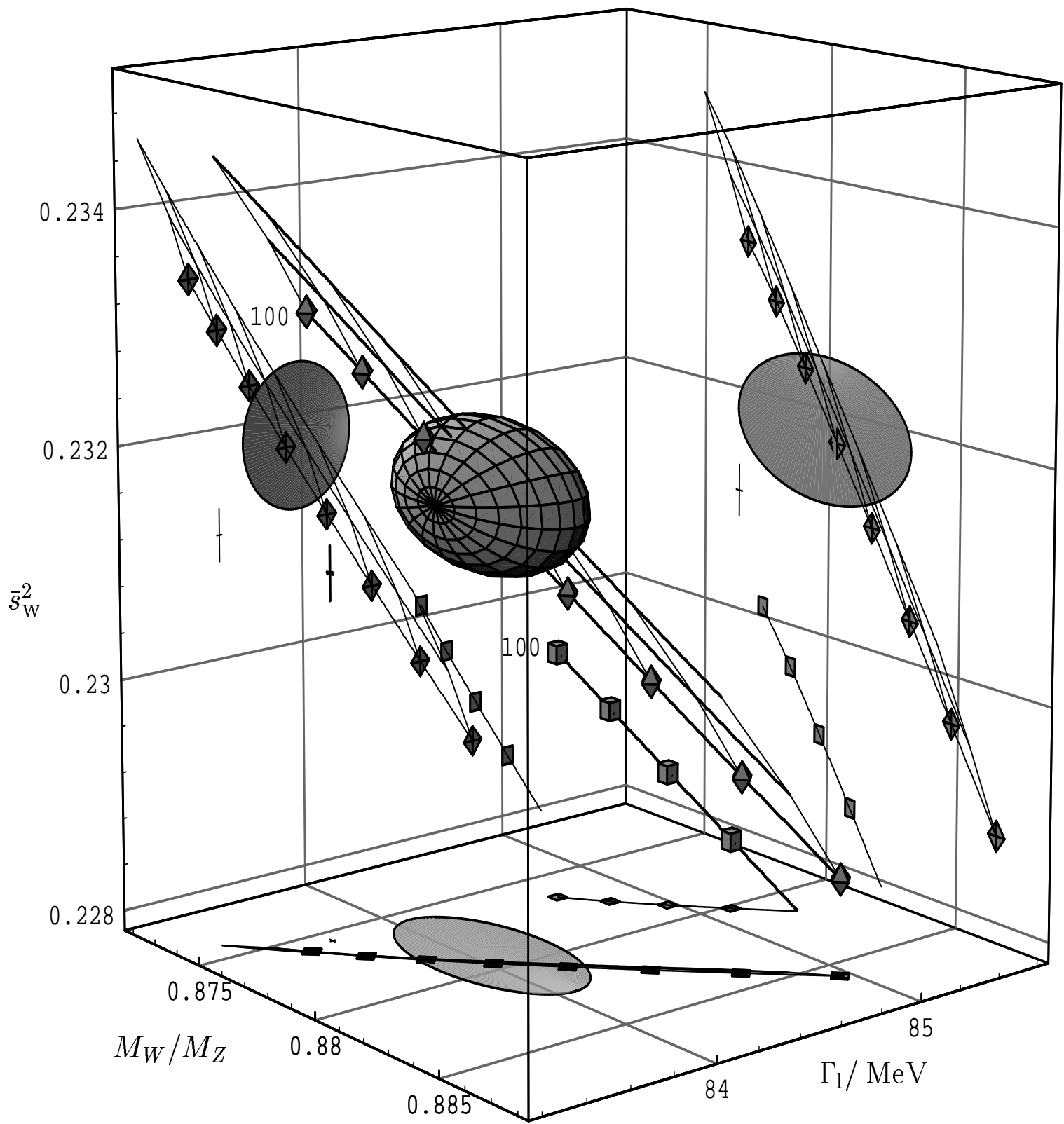
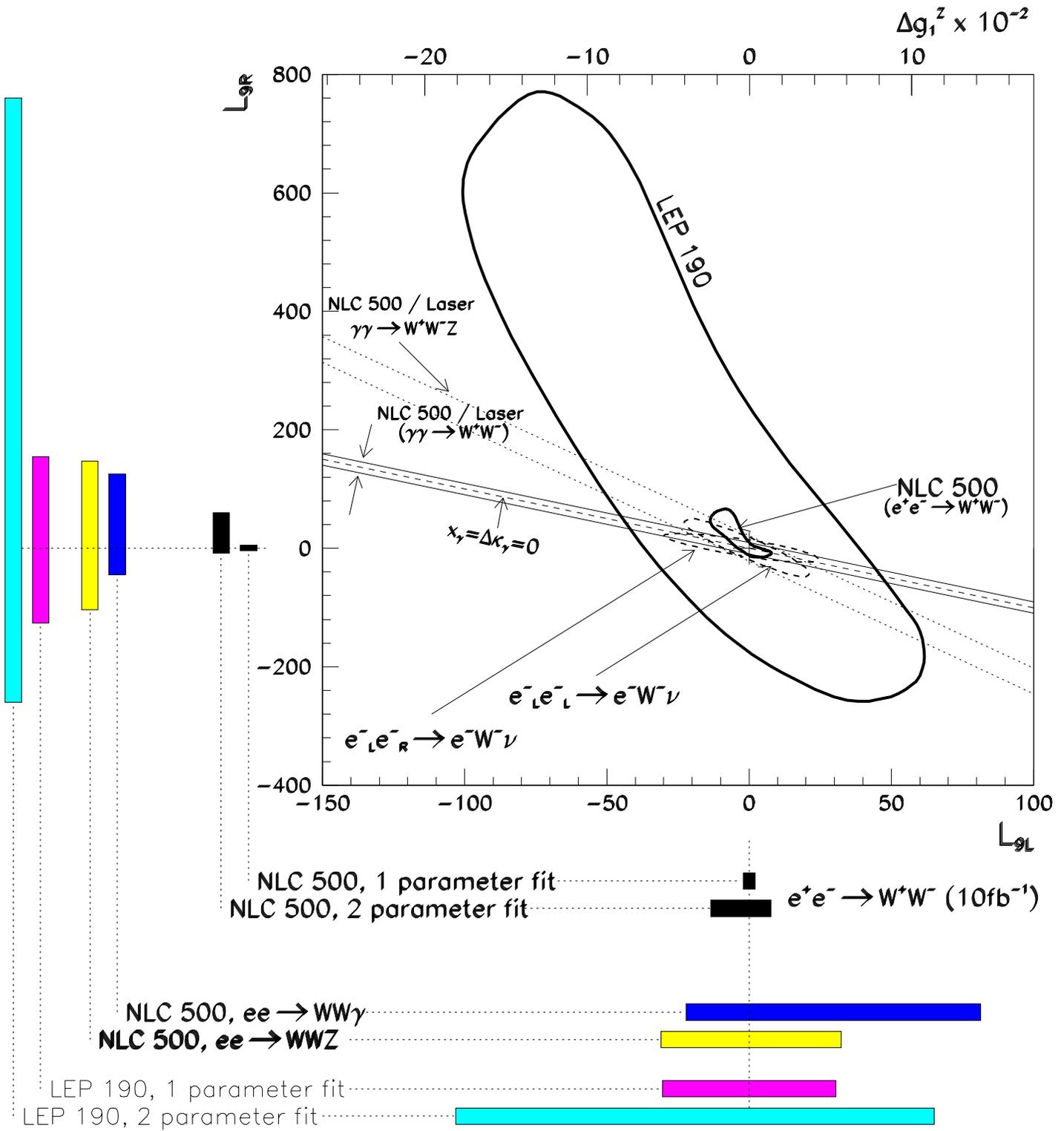
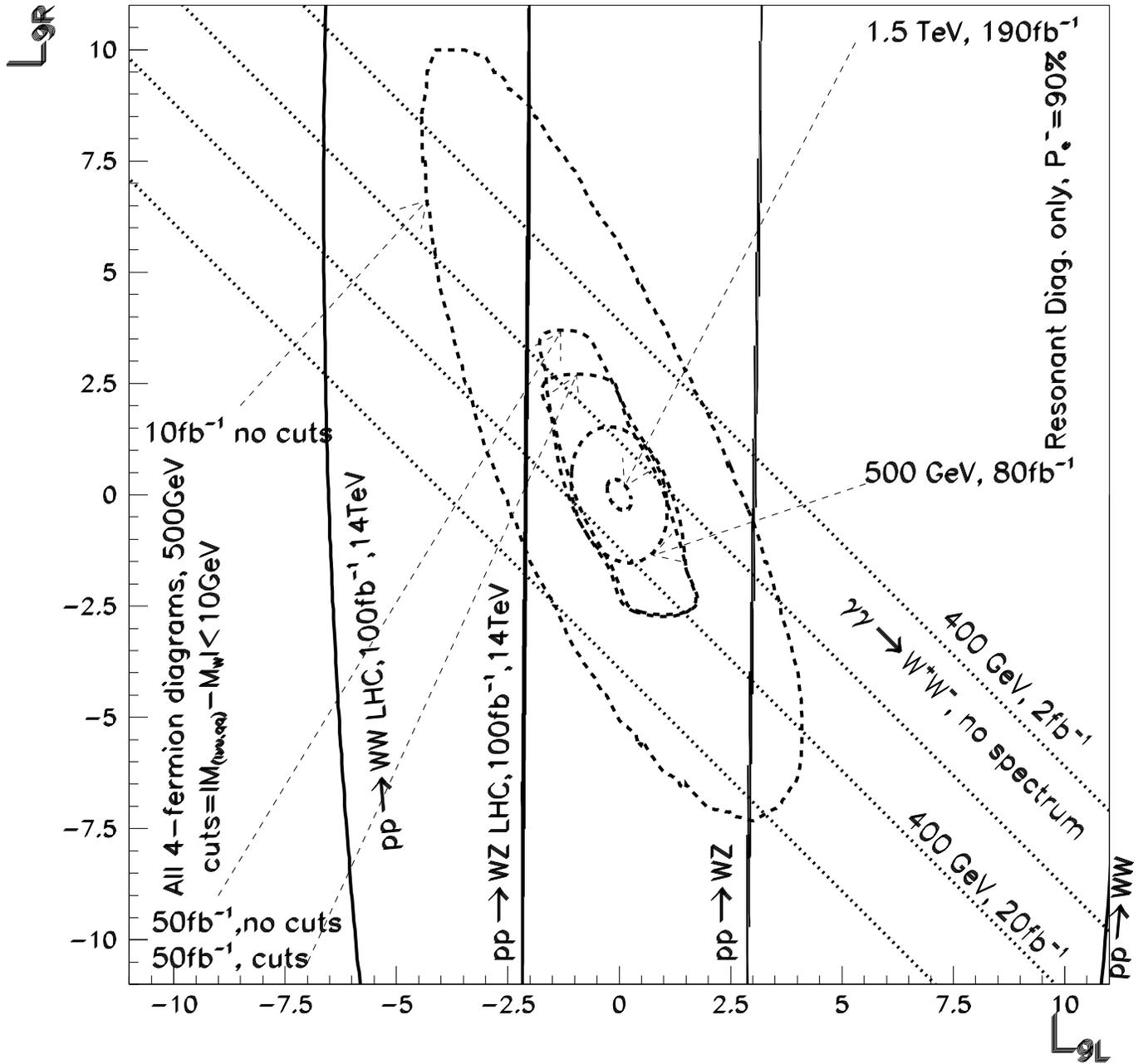
\end{document}